\documentclass[12pt, epsf]{article}  
\usepackage{epsf}
\begin{document}
\newcommand{\sheptitle}
{Degenerate and Inverted Hierarchical Models of Majorana Neutrinos 
from See-saw Mechanism}
\newcommand{\shepauthor}
{N. Nimai Singh${^\dag}$  and Mahadev Patgiri${^\ddag}$}
\newcommand{\shepaddress}
{${^\dag}$ Department of Physics, Gauhati University, Guwahati-781014, India\\
${\ddag}$Dpartment of Physics, Cotton College, Guwahati-781001, India}
\newcommand{\shepabstract}
{In order to explain  the recently reported  result on 
the double beta decay $(0\nu\beta\beta)$ experiment, together with the earlier 
established
data on LMA MSW solar and atmospheric neutrino oscillations, we generate the
textures of the nearly degenerate as well as the inverted
hierarchical models of the left-handed Majorana neutrino mass matrices 
within the framework of the see-saw mechanism  in a model independent way.
 The leptonic mixings are  generated from the texture of the right-
handed Majorana mass matrices while keeping the Dirac neutrino mass matrices 
 in diagonal forms.\\
PACS numbers: 12.10.Dm;12.15.Ff;14.60.Gh\\
Keywords: Majorana neutrino mass, degenerate, inverted hierarchical, mixing angles,
mass eigenvalues.}
\begin{titlepage}
\begin{flushright}
hep-ph/0204021
\end{flushright}
\begin{center}
{\large{\bf\sheptitle}}
\bigskip\\
\shepauthor
\\
\mbox{}\\
{\it\shepaddress}\\
\vspace{.5in}
{\bf Abstract}
\bigskip
\end{center}
\setcounter{page}{0}
\shepabstract
\end{titlepage}
\section{Introduction}
 The recently reported experimental result on double beta decay ($0\nu\beta\beta$) 
shows the possible evidence for the non-zero Majorana mass of the electron neutrino
in  the range of $m_{ee}= (0.05- 0.86)$ eV at 95\% C.L. with the best value of
 $m_{ee}= 0.4$ eV[1]. There are  certain 
important implications of this result. Firstly, it rules out all models predicting the 
Dirac neutrino masses, leaving only the option for the Majorana type of neutrinos. 
It also allows the lepton number
violating processes such as leptogenesis in a natural way. Secondly, this result together
with the earlier experimental data on the atmospheric[2] and the solar[3] neutrino 
oscillations, allows
only the degenerate and the inverted hierarchical solutions for the three generation  
left-handed Majorana neutrinos[1,4].

 In the above context, it is important to construct theoretical models which can 
predict the degenerate and inverted hierarchical patterns of the Majorana 
neutrino mass matrices
within  the framework of the grand unified theories(GUTs) with or without 
supersymmetry[4,5]. In this short paper we attempt
to generate the degenerate as well as the inverted hierarchical pattern of the left-handed 
Majorana neutrino mass matrices using the see-saw formula in a model independent way.
This is, in fact, a continuation of our earlier work[6]  where the neutrino mixings 
are provided from the texture of the right-handed Majorana mass matrix $M_{RR}$,
while keeping the Dirac neutrino mass matrix $m_{LR}$ in the diagonal form. 
We had taken the Dirac neutrino mass matrix $m_{LR}$ as either the charged lepton mass matrix
( $m_{LR}$ =$ tan\beta$ $m_{l}$ referred to as case(i)) or the up-quark mass matrix
( $m_{LR} = m_{up}$ referred to as case(ii))[7]. While referring to the earlier paper[6]
for details, the model successfully generated both the hierarchical and the inverted
hierarchical (having opposite sign mass eigenvalues) neutrino mass matrices as a
result of the proper choice of the parameters in texture of $M_{RR}$. In section 2 we present 
the generation of the degenerate as well as inverted hierarchical neutrino mass matrices using 
the seesaw formula, and their predictions on mass eigenvalues and mixing angles. Section 3 is devoted to 
summary and conclusion.

\section{Neutrino mass matrices from See-saw formula}

 The left-handed Majorana neutrino mass matrix $m_{LL}$ is
given by the celebrated see-saw formula[8],
\begin{equation}
m_{LL} = -  m_{LR} M^{-1}_{RR} m^{T}_{LR}
\end{equation}
where  $m_{LR}$ is the Dirac neutrino mass matrix in the left-right (LR) convention[9]. 
  The leptonic (MNS) mixing matrix is now given by $V_{MNS} = V^{\dag}_{\nu L}$
where $m^{diag}_{LL}$ = $V_{\nu L}$ $ m_{LL}$ $V^{T}_{\nu L}$. Here both  $m_{LR}$
and the charged lepton mass matrix $m_{l}$ are taken as diagonal, whereas the
right-handed Majorana neutrino mass matrix $M_{RR}$ as non-diagonal. Using 
the see-saw formula (1) we generate both patterns of $m_{LL}$ viz.,
(I) nearly degenerate and (II) inverted hierarchical neutrino mass matrices 
with same sign mass eigenvalues. We conentrate here only on the cases which have  bimaximal mixings with (+) entries in $m_{LL}$  listed in Table-I.\\
\\

Table-I: Zeroth order neutrino mass matrices with texture zeros allowed by
the result on $0\nu\beta\beta$ decay. The (+) entries in the matrices correspond 
to the LMA MSW solution with bimaximal mixings whereas (-) correspond to
SMA MSW solution with single maximal mixing[4,10].

\begin{center}
\begin{tabular}{ccc} \hline \hline
\it{Type} & \it{$m_{LL}$} & \it{$m^{diag}_{LL}$}\\
\hline
I(A) & $\left( \begin{array}{ccc}
 1 & 0 &  0\\
 0 & 0 & \pm 1\\
 0& \pm 1 & 0
\end{array}
\right) m_{0}$ & $Diag( 1, \pm 1, \mp 1) m_{0}$\\

I(B) & $\left( \begin{array}{ccc}
\pm 1 & 0 &  0\\
 0 & 1 & 0\\
 0 & 0 & 1
\end{array}
\right) m_{0}$ & $Diag(\pm 1, 1, 1) m_{0}$\\

II(A) & $\left( \begin{array}{ccc}
 1 & 0 &  0\\
 0 & \pm \frac{1}{2}& \pm \frac{1}{2}\\
 0& \pm \frac{1}{2} & \pm \frac{1}{2}
\end{array}
\right) m_{0}$ & $Diag( 1, \pm 1, 0) m_{0}$\\

II(B) &  $\left( \begin{array}{ccc}
 0 & 1 &  1\\
 1 & 0 & 0\\
 1 & 0 & 0
\end{array}
\right) m_{0}$ & $Diag( 1, -1, 0) m_{0}$\\
\hline\hline
\end{tabular}
\end{center}

 {\bf I(A). Nearly degenerate mass matrix with opposite sign mass
eigenvalues:}\\

 The most common texture for  the nearly degenerate mass matrix 
$m_{LL}$ with opposite mass eigenvalues is represented by[5] 
\begin{equation}
m_{LL}=
\left( \begin{array}{ccc}
(1-2\delta_{1}-2\delta_{2}) & -\delta_{1} &  -\delta_{1}\\
 -\delta_{1} & -\delta_{2} & (1-\delta_{2})\\
 -\delta_{1}  & (1-\delta_{2}) &  -\delta_{2}
\end{array}
\right) m_{0}
\end{equation}
where $m_{0}$ controls the overall magnitude of the masses of the 
neutrinos whereas $\delta_{1}$ and $\delta_{2}$ give the desired
splittings for solar and atmospheric data. When $\delta_{1}=\delta_{2}= 0$, 
Eq.(2) reduces to the zeroth order mass matrix of  
the Type I(A) in Table-I, with no splittings[5,10].

The diagonalisation of $m_{LL}$ in Eq.(2) leads to the following
eigenvalues and mixing angles [see Appendix]:\\
$m_{\nu_{1}}\simeq (1- 2\delta_{2}- (\sqrt{3}+1)\delta_{1})m_{0},$\\
$m_{\nu_{2}}\simeq (1- 2\delta_{2}+ (\sqrt{3}-1)\delta_{1})m_{0},$\\
$m_{\nu_{3}}\simeq -m_{0},$\\
$\sin^{2}2\theta_{12}= \frac{2}{3}$, $\sin^{2}2\theta_{23}= 1$,
$\sin^{2}2\theta_{13}= 0.$
\\
For the choice[5] of the values of the parameters $m_{0}=0.4$ eV,
$\delta_{1}=3.6 \times 10^{-5}$, $\delta_{2}=3.9 \times 10^{-3}$,
Eq.(2) leads to the following numerical  predictions \\
\underline{\it{Mixing angles}}:\\
$sin^{2}2\theta_{12}= 0.67$, $sin^{2}2\theta_{23} \approx 1.0$,
$|V_{e3}| = 1.5 \times 10^{-14}$,\\

\underline{\it{Mass eigenvalues}}:\\
$m_{\nu_{i}} = ( 0.39684, 0.396892, -0.4)$ eV, $i = 1, 2, 3$;
$\Delta m^{2}_{12} = 4.13 \times  10^{-5} eV^{2}$ and
$\Delta m^{2}_{23} = 2.48 \times 10^{-3} eV^{2}$.\\

The prediction on solar mixing angle  is consistent with the LMA MSW solution[3].\\

\underline{Case (i) where $m_{LR}$ =$tan\beta$ $m_{l}$}[6,7]\\

 The degenerate mass matrix $m_{LL}$ in Eq.(2) is now 
generated through see-saw formula[8] in Eq.(1)
for the following choices of $m_{LR}$ and $M_{RR}$ respectively:
\begin{equation}
m_{LR}= tan\beta
\left( \begin{array}{ccc}
\lambda^{6} & 0 & 0 \\
0 & \lambda^{2} & 0 \\
0 & 0 & 1
\end{array}
\right) m_{\tau},
\end{equation}
and
\begin{equation}
M_{RR} =
\left( \begin{array}{ccc}
(1+2\delta_{1}+2\delta_{2})\lambda^{12} & \delta_{1}\lambda^{8} &  \delta_{1}\lambda^{6}\\
 \delta_{1}\lambda^{8} &  \delta_{2}\lambda^{4} &  (1+\delta_{2})\lambda^{2}\\
 \delta_{1}\lambda^{6} &  (1+\delta_{2})\lambda^{2} & \delta_{2}
\end{array}
\right) v_{R}
\end{equation}    
The inverse of $M_{RR}$ has a simple form for $\delta_{2}>\delta_{1}$ and $\delta_{1},\delta_{2}<<1$,  
$$
M_{RR}^{-1} =
\left( \begin{array}{ccc}
(1-2\delta_{1}-2\delta_{2})\lambda^{-12} & -\delta_{1}\lambda^{-8} &  -\delta_{1}\lambda^{-6}\\
 -\delta_{1}\lambda^{-8} &  -\delta_{2}\lambda^{-4} &  (1-\delta_{2})\lambda^{-2}\\
 -\delta_{1}\lambda^{-6} &  (1-\delta_{2})\lambda^{-2} & -\delta_{2}
\end{array}
\right) (v_{R}^{-1})
$$    
The expression for $m_{0}$ in Eq.(2) is worked out as
$m_{0} = m^{2}_{\tau} tan^{2}{\beta}/ v_{R}$. 
For input values of $m_{0} = 0.4 eV$, $tan\beta = 40$, $m_{\tau} = 1.7$ GeV,
and $\lambda = 0.22$, the see-saw scale is calculated as $v_{R} \approx 10^{13}$ GeV.  
This in turn  gives the masses of the three right-handed Majorana neutrinos after the 
diagonalisation of $M_{RR}$:\\
$|M^{diag}_{RR}| = ( 4.67\times 10^{11}, 1.296\times10^{5}, 5.06\times 10^{4})$ GeV.
\\

\underline{Case(ii) where $m_{LR} = m_{up}$}[6,7]\\

 The texture of $m_{LL}$ in Eq.(2) is again realised through the see-saw formula (1)
with the following textures of $m_{LR}$ and $M_{RR}$:
\begin{equation}
m_{LR}= 
\left( \begin{array}{ccc}
\lambda^{8} & 0 & 0 \\
0 & \lambda^{4} & 0 \\
0 & 0 & 1
\end{array}
\right) m_{t},
\end{equation}
and
\begin{equation}
M_{RR} =
\left( \begin{array}{ccc}
(1+2\delta_{1}+2\delta_{2})\lambda^{16} & \delta_{1}\lambda^{12} &  \delta_{1}\lambda^{8}\\
 \delta_{1}\lambda^{12} &  \delta_{2}\lambda^{8} &  (1+\delta_{2})\lambda^{4}\\
 \delta_{1}\lambda^{8} &  (1+\delta_{2})\lambda^{4} & \delta_{2}
\end{array}
\right) v_{R}
\end{equation}

We have $m_{0} = m^{2}_{t}/ v_{R}$ in Eq.(2), and with the input values $m_{0} = 0.4$ eV,
$m_{t} = 200$ GeV, we obtain $v_{R}= 10^{14}$ GeV and the mass eigenvalues of the right-handed
Majorana neutrinos: 
$|M^{diag}_{RR}| = ( 1.105 \times 10^{11}, 3.035 \times 10^{3}, 5.005 \times 10^{11})$ GeV.\\

{\bf I(B). Nearly degenerate mass matrix with same sign mass eigenvalues:}\\

We propose another form of the nearly degenerate mass matrix,
\begin{equation}
m_{LL}=
\left( \begin{array}{ccc}
(1-2\delta_{1}-2\delta_{2}) & -\delta_{1} &  -\delta_{1}\\
 -\delta_{1} &(1 -\delta_{2}) & -\delta_{2}\\
 -\delta_{1}  & -\delta_{2} &  (1-\delta_{2})
\end{array}
\right) m_{0}
\end{equation}

The diagonalisation of $m_{LL}$ in Eq.(7) leads to [see Appendix]\\
$m_{\nu_{1}}\simeq (1- 2\delta_{2}- (\sqrt{3}+1)\delta_{1})m_{0},$\\
$m_{\nu_{2}}\simeq (1- 2\delta_{2}+ (\sqrt{3}-1)\delta_{1})m_{0},$\\
$m_{\nu_{3}}\simeq m_{0}$,\\
$\sin^{2}2\theta_{12}= \frac{2}{3}$, $\sin^{2}2\theta_{23}= 1$,
$\sin^{2}2\theta_{13}= 0$.\\

The numerical solution leads to  
$m_{\nu_{i}} = (0.39684, 0.396892, 0.4)$ eV, $i=1, 2, 3$
 for the same choices of the input values of $\delta_{1,2}$ and $m_{0}$
as in Eq.(2). Further, the predictions on the three mixing angles
are the same as in Eq.(2). 
When $\delta_{1}= \delta_{2}= 0$, it reduces to the type I(B)
in the Table-I.\\

 This form of the mass matrix(7) can be realised
in the see-saw mechanism(1) using the following textures of $m_{LR}$ and $M_{RR}$:\\

\underline{Case(i) where $m_{LR}$ = $\tan\beta$ $m_{l}$}\\

Here $m_{LR}$ is  given in Eq.(3)
and the right-handed neutrino mass matrix takes the form 
\begin{equation}
M_{RR} =
\left( \begin{array}{ccc}
(1+2\delta_{1}+2\delta_{2})\lambda^{12} & \delta_{1}\lambda^{8} &  \delta_{1}\lambda^{6}\\
 \delta_{1}\lambda^{8} &  (1+\delta_{2})\lambda^{4} &  \delta_{2}\lambda^{2}\\
 \delta_{1}\lambda^{6} &  \delta_{2}\lambda^{2} & (1+\delta_{2})
\end{array}
\right) v_{R}
\end{equation}
The inverse of $M_{RR}$ has a simple form for $\delta_{2}>\delta_{1}$ and $\delta_{1},\delta_{2}<<1$,  
$$
M_{RR}^{-1} =
\left( \begin{array}{ccc}
(1-2\delta_{1}-2\delta_{2})\lambda^{-12} & -\delta_{1}\lambda^{-8} &  -\delta_{1}\lambda^{-6}\\
 -\delta_{1}\lambda^{-8} & (1 -\delta_{2})\lambda^{-4} &  \delta_{2}\lambda^{-2}\\
 -\delta_{1}\lambda^{-6} &  -\delta_{2}\lambda^{-2} & (1-\delta_{2})
\end{array}
\right) (v_{R}^{-1})
$$    
The expression for $m_{0}$ in Eq.(7) is again worked out as
$m_{0} = m^{2}_{\tau} tan^{2}{\beta}/ v_{R}$.\\

\underline{Case(ii) where $m_{LR}=m_{up}$}\\

The texture of $m_{LR}$ is  given in Eq.(5), and the texture of $M_{RR}$ has the form  
\begin{equation}
M_{RR} =
\left( \begin{array}{ccc}
(1+2\delta_{1}+2\delta_{2})\lambda^{16} & \delta_{1}\lambda^{12} &  \delta_{1}\lambda^{8}\\
 \delta_{1}\lambda^{12} &  (1+\delta_{2})\lambda^{8} &  \delta_{2}\lambda^{4}\\
 \delta_{1}\lambda^{8} &  \delta_{2}\lambda^{4} & (1+\delta_{2})
\end{array}
\right) v_{R}
\end{equation}  
and $m_{0} = m^{2}_{t}/ v_{R}$.\\

{\bf II(A). Inverted hierarchical mass matrix with same sign mass eigenvalues:}\\

 The most general form of the inverted hierarchical mass matrix $m_{LL}$ 
with the same sign mass eigenvalues can be expressed  as
\begin{equation}
m_{LL}=
\left( \begin{array}{ccc}
(1-2\epsilon) & -\epsilon &  -\epsilon\\
 -\epsilon & a  & (a-\eta)\\
 -\epsilon  & (a-\eta) & a  
\end{array}
\right) m'_{0}
\end{equation}
where $a =0.5$ and $m'_{0}$ is the overall factor for 
the masses of the neutrinos.
The parameters $\epsilon$ and $\eta$ give the desired splittings for solar
and atmospheric data.\\

 The diagonalisation of $m_{LL}$ in Eq.(10)
leads to the following eigenvalues and mixing angles [see Appendix]:\\
$m_{\nu_{1}}\simeq (1- (\sqrt{3}+1)\epsilon-
\frac{\eta}{2}+\frac{\sqrt{\eta\epsilon}}{6})m'_{0},$\\
$m_{\nu_{2}}\simeq (1+ (\sqrt{3}-1)\epsilon
-\frac{\eta}{2}-\frac{\sqrt{\eta\epsilon}}{6})m'_{0},$\\
$m_{\nu_{3}}$$\simeq$ $\eta$ $m'_{0},$\\
and mixing angles:\\
 $\sin^{2}2\theta_{12}= \frac{2}{3}$, 
$\sin^{2}2\theta_{23}= 1$,
$\sin^{2}2\theta_{13}= 0.$\\
When $\epsilon = \eta = 0$, Eq.(10) reduces to
the zeroth order mass matrix of the type II(A) in Table-I,
with no solar splitting[4,10].  
For solution of the 
LMA MSW solar data and atmospheric neutrino oscillation, we have the choice
of the parameters $m'_{0} = 0.05$ eV, $\epsilon = 0.002$ and $\eta = 0.0001$
leading to the following predictions:\\

\underline{\it{Mixing angles}:}\\
 $sin^{2}2\theta_{12}= 0.67$, $sin^{2}2\theta_{23} \approx 1.0$,
$|V_{e3}| = 3.04 \times 10^{-13}$,\\

\underline{\it{Mass eigenvalues}}:\\
$m_{\nu_{i}} = ( 0.05007, 0.04973, 0.000005)$ eV, $i = 1, 2, 3$; leading to
$\Delta m^{2}_{12} = 3.393 \times  10^{-5} eV^{2}$ and 
$\Delta m^{2}_{23} = 2.47 \times 10^{-3} eV^{2}$.

Here the neutrino mass eigenvalues are of the same sign and this differs from 
the other inverted hierarchical mass matrix having the following form[11],
\begin{equation}
m_{LL}=
\left( \begin{array}{ccc}
\epsilon  & 1 &  1\\
 1 & \delta_{1}  & \delta_{2}\\
 1  & \delta_{2} & \delta_{1}  
\end{array}
\right) m'_{0},  \epsilon,\delta_{1},\delta_{2}<< 1
\end{equation}
which gives opposite sign mass eigenvalues. For $ \delta_{1}$, $\delta_{2}$,
$\epsilon$ = 0, it leads to the type II(B) in Table-I. 
This structure has been successfully generated
within this model[6], and the radiative correction has been  found to be weak[12,13].
We shall not address again this model  here. Instead, we concentrate on 
the generation of the texture of  $m_{LL}$ given in Eq.(10) from the see-saw mechanism (1).\\

\underline{Case(i) when $m_{LR}$ = $tan\beta$ $m_{l}$}\\

 The inverted hierarchical mass mtrix $m_{LL}$ in Eq.(10) 
can now be realised with the choice of $m_{LR}$ = $tan\beta$ $m_{l}$
given in Eq.(3) and $M_{RR}$ of the following form
\begin{equation}
M_{RR}=
\left( \begin{array} {ccc}
2a\eta (1+2\epsilon) \lambda^{12} & \eta \epsilon \lambda^{8} & \eta \epsilon \lambda^{6}\\ 
\eta \epsilon \lambda^{8}  &  a\lambda^{4}  &  -(a-\eta)\lambda^{2}\\
\eta \epsilon \lambda^{6}  &  -(a-\eta)\lambda^{2} & a
\end{array} \right) \frac{v_{R}}{2a\eta}
\end{equation}
where $M_{RR}^{-1}$ has a simple form for $\epsilon>\eta$ 
and $\epsilon, \eta <<1$,
$$
M_{RR}^{-1}=
\left( \begin{array} {ccc}
 (1-2\epsilon) \lambda^{-12} & - \epsilon \lambda^{-8} & - \epsilon \lambda^{-6}\\ 
- \epsilon \lambda^{-8}  &  a\lambda^{-4}  &  (a-\eta)\lambda^{-2}\\
- \epsilon \lambda^{-6}  &  (a-\eta)\lambda^{-2} & a
\end{array} \right)(v_{R}^{-1})
$$
The expression for $m'_{0}$ in Eq.(10) is given by
$m'_{0} = m^{2}_{\tau} tan^{2}\beta/v_{R}$. For input values of $m'_{0}= 0.05$ eV, $tan\beta= 5$,
$m_{\tau}= 1.7$ GeV, we obtain $v_{R}= 1.445\times 10^{12}$ GeV which 
leads to $|M^{diag}_{RR}|= (9.742\times 10^{3}, 2.831
\times 10^{4}, 7.24\times10^{15})$ GeV.\\ 

\underline{Case(ii) when $m_{LR}= m_{up}$}\\
 The texture of $m_{LL}$ in Eq.(10) is again realised using   $m_{LR}= m_{up}$ given in
Eq.(5) and the texture of $M_{RR}$:
\begin{equation}
M_{RR}=
\left( \begin{array} {ccc}
2a\eta (1+2\epsilon) \lambda^{16} & \eta \epsilon \lambda^{12} & \eta \epsilon \lambda^{8}\\ 
\eta \epsilon \lambda^{12}  &  a\lambda^{8}  &  -(a-\eta)\lambda^{4}\\
\eta \epsilon \lambda^{8}  &  -(a-\eta)\lambda^{4} & a
\end{array} \right) \frac{v_{R}}{2a\eta}
\end{equation}
which leads to $m'_{0}= m^{2}_{t}/v_{R}$ in Eq.(10). Using the input values
$m'_{0}=0.05 eV$, $m_{t}= 200$GeV, we have $v_{R}= 8\times 10^{14}$ GeV and
$|M^{diag}_{RR}|= ( 2.4\times 10^{4}, 4\times 10^{18}, 2.4\times 10^{9})$ GeV
where the mass of the heaviest right-handed Majorana neutrino lies above
the GUT scale but below the Planck scale[10].

In the above calculation we have confined to the (+) entries of $m_{LL}$ in Table-I 
and these are consistent  
with the LMA MSW solution of the solar neutrino problem.
 The (-) entries in the texture of $m_{LL}$ in Table-I which generally 
lead  to SMA MSW solar neutrino oscillation solution, can also be worked out 
 in this model under the same framework with the proper choices of ($\pm$) signs before  the
elements of the  $M_{RR}$.
 
A few comments on the stability condition under radiative 
corrections are in order. The nearly degenerate mass matrices
$m_{LL}$ in Eqs.(2) and (7) are found to be unstable under 
radiative corection in minimal supersymmetric standard model.
The inverted hierarchical mass matrix with the same mass eigenvalues
given in Eq.(10) is also found to be unstable under radiative
correction. However, the inverted hierarchical mass matrix given in
Eq.(11) with opposite sign mass eigenvalues is stable under
radiative corection [12,13].  

\section{Summary}
 In summary, we generate the textures of the nearly degenerate 
as well as the inverted hierarchical left-handed Majorana neutrino 
mass matrices from the see-saw formula using the diagonal form of 
the Dirac mass matrix and non-diagonal form of the right-handed Majorana
neutrino mass matrix. The predictions on lepton mixing angles 
$sin^{2}2\theta_{12}\approx 0.67$, $sin^{2}2\theta_{23}\approx 1.0$ and
$|V_{e3}|\approx 0$ are in excellent agreement with the experimental 
values. This is also true for the predictions of  $\Delta m^{2}_{12}$ 
and $\Delta m^{2}_{23}$ which are necessary for the 
$0\nu \beta \beta$ decays, LMA MSW solar oscillation
and atmospheric oscillation data. In all cases the masses of the right-handed Majorana
neutrinos are  above the weak scale. An interesting observation is that the
prediction on the solar mixing angle is large but not maximal without any extra fine-tuning.
Though the present analysis is a model independent one, it may be a useful
guide for building models under the framework of grand unified theories
with the extended flavour U(1) symmetry.\\

\section *{Appendix}
\underline{Diagonalisation of the mass matrix $m_{LL}$}\\      

Although the diagonalisation of $m_{LL}$ is trivial[5], we find it convenient 
to follow the general result given in Ref.[15]. The  neutrino mass matrix $m_{LL}$ of the general form 
$$m_{LL} = \left(\begin{array}{ccc}
a & b & c\\ b & d & e\\ c & e & f
\end{array}\right)$$
where $c= -t_{23} b$ and $t_{23}= \sin\theta_{23}/\cos\theta_{23}$,
$f= d + (t^{-1}_{23}- t_{23}) e$ \\
can be diagonalised by $V_{MNS}$ 
$$ V_{MNS} = \left(\begin{array}{ccc}
\cos\theta_{12} & \sin\theta_{12} & 0\\
-\cos\theta_{23} \sin\theta_{12} & \cos\theta_{23} 
\sin\theta_{12} & \sin\theta_{23}\\
\sin\theta_{23} \sin\theta_{12} & -\sin\theta_{23} \cos\theta_{12}
& \cos\theta_{23}
\end{array}\right)$$
where we have taken $\sin\theta_{13}=0$. 
This mixing MNS matrix  transforms $|\nu_{i}>$ with the masses ($m_{\nu_{1}}$,
$m_{\nu_{2}}$,$m_{\nu_{3}}$) into $|\nu_{f}>$ via\\
$|\nu_{f}> = V_{MNS} |\nu_{i}>$, $f= e, \mu, \tau$ and $i=1, 2, 3.$
The mass eigenvalues  and $\theta_{12}$ are calculated as \\
$m_{\nu 1} = a - \frac{1}{2}\sqrt{\frac{b^{2}+c^{2}}{2}} 
(x + \eta \sqrt{x^{2}+8}),$ $m_{\nu 2}$ = 
($ \eta$$\rightarrow$$ -\eta$ in $m_{\nu 1}$),\\
$m_{\nu 3} = d + t^{-2}_{23}( d - a + x \sqrt{\frac{b^{2}+ c^{2}}{2}})$,
$\sin^{2}2\theta_{12} = \frac{8}{8+x^{2}}$\\
 with $x = (a- d + t_{23} e)/(\sqrt{\frac{b^{2} + c^{2}}{2}})$
where $|m_{\nu 1}| < |m_{\nu 2}|$ is always maintained by adjusting the sign
of $\eta( = \pm 1)$.\\   


\begin{thebibliography}{99}
\bibitem{ref1}H.V.Klapdor-Kleingrothaus, A.Dietz, H.L.Harney, I.V.Krivosheina, Mod. Phys. Lett. 
{\bf A16}(2001)2409.
\bibitem{ref2}Y.Fukuda et al, Super-Kamiokande Collaboration, Phys. Rev. Lett. {\bf 85}(2000)3999. 
\bibitem{ref3}Y.Fukuda et al, Super-Kamiokande Collaboration, Phys. Rev. Lett. {\bf 86}(2001)5656
 and references therein. See also Q.R.Ahmed et al, SNO Collaboration, Phys. Rev. Lett. {\bf 87}(2001)
071301.
\bibitem{ref4}H.V.Klapdor-Kleingrothaus, U.Sarkar, {\bf hep-ph/0201224}.
\bibitem{ref5}E.Ma, {\bf hep-ph/0201225}
\bibitem{ref6}N.Nimai Singh, Mahadev Patgiri,{\bf hep-ph/0111319}
\bibitem{ref7}K.S.Babu, B.Dutta, R.N.Mohapatra, Phys.Lett. {\bf B458}(1999)93, {\bf hep-ph/9904366}.
\bibitem{ref8}M.Gell-Mann, P.Ramond, R.Slansky, in: Supergravity, North-Holland, Amsterdam, 1979;
 T.Yanagida, in : Proc. of the workshop on Unified Theory and Baryon number of the Universe, KEK, 
 Japan, 1979; R.N. Mohapatra, G.Senjanovic, Phys. Rev. Lett. {\bf 44}(1980)912.
\bibitem{ref9}S.F.King, N.Nimai Singh, Nucl.Phys. {\bf B591}(2000)3; Nucl.Phys. {\bf B596}(2001)81.
\bibitem{ref10}G.Altarelli, F.Feruglio, Phys.Rept. {\bf 320}(1999)295.
\bibitem{ref11}An incomplete list: R.Barbieri, L.Hall, D.Smith, A.Strumia,  N.Weiner, JHEP {\bf 9812}
(1998)17, {\bf hep-ph/9807235}; A.S.Joshipura, S.D.Rindani, Eur.Phys. J.{\bf C14}(2000)85; Q.Shafi, Z.
Tavartkiladze, Phys. Lett. {\bf B482}(2000)145; S.F.King, N.Nimai Singh, Nucl.Phys.{\bf B596}(2001)81;
M.Patgiri, N.Nimai Singh, {\bf hep-ph/0112123}; K.S.Babu, R.N.Mohapatra, {\bf hep-ph/0201176}.
\bibitem{ref12}P.H.Chankowski, W.Krolikowski, S.Pokorski, Phys.Lett. {\bf B473}(2000)109; 
N.Haba, N.Okamura, Eur. Phys. J.{\bf C14}(2000)347.
\bibitem{ref13}S.F.King, N.Nimai Singh. Nucl.Phys.{\bf B596}(2001)81 ;
M.Patgiri, N.Nimai Singh, {\bf hep-ph/0112123}.
\bibitem{ref14}H.V.Klapdor-Kleingrothaus, U.Sarkar, Mod. Phys. Lett. {\bf A16}(2001)2469.
\bibitem{ref15}T.Kitabayashi, M.Yasue, {\bf hep-ph/0112287}.

\end{thebibliography}
                                                          \end{document}